\documentclass[preprint,preprintnumbers,amsmath,amssymb,superscriptaddress]{revtex4}

\pdfoutput=1

\def \dedp {\textstyle \frac{\partial e}{\partial p} }
 
\newcommand{\be}{\begin{equation}}
\newcommand{\ee}{\end{equation}}
\newcommand{\ba}{\begin{eqnarray}}
\newcommand{\ea}{\end{eqnarray}}
\newcommand{\bd}{\begin{displaymath}}
\newcommand{\ed}{\end{displaymath}}

\newcommand{\commentout}[1]{{}}

\usepackage{amssymb}
\usepackage{color}
\usepackage{slashed}
\usepackage{amsmath}
\usepackage{graphicx}
\usepackage[caption=false]{subfig}
\usepackage{epstopdf}
\usepackage{dcolumn}
\usepackage{bm}
\usepackage{hyperref}

\begin{document}

\title{\bf{Thermally Fluctuating Second-Order Viscous Hydrodynamics and Heavy-Ion Collisions}}
\author{C. Young}
\email{young@physics.umn.edu}
\affiliation{School of Physics \& Astronomy, University of Minnesota, Minneapolis, MN 55455, U.S.A.}
\author{J. I. Kapusta}
\email{kapusta@physics.umn.edu}
\affiliation{School of Physics \& Astronomy, University of Minnesota, Minneapolis, MN 55455, U.S.A.}
\author{C. Gale}
\email{gale@physics.mcgill.ca}
\affiliation{Department of Physics, McGill University, 3600 rue University Montr\'eal QC H3A 2T8 Canada}
\author{S. Jeon}
\email{jeon@physics.mcgill.ca}
\affiliation{Department of Physics, McGill University, 3600 rue University Montr\'eal QC H3A 2T8 Canada}
\author{B. Schenke}
\email{bschenke@quark.phy.bnl.gov}
\affiliation{Physics Dept, Bldg. 510A, Brookhaven National Laboratory, Upton, NY 11973}
\date{\today}

\begin{abstract}

The fluctuation-dissipation theorem requires the presence of thermal noise in viscous fluids. The time and length scales of heavy ion collisions are small enough so that the thermal noise can have a measurable effect on observables.  Thermal noise is included in numerical simulations of high energy lead-lead collisions, increasing average values of the momentum eccentricity and contributing to its event by event fluctuations.

\end{abstract}

\maketitle

\section{Introduction}
\label{introduction}

Observables in high energy heavy ion collisions are consistent with the formation of a very hot, nearly ideal fluid \cite{Adcox:2004mh, Adams:2005dq}. It remains difficult to understand the physical reasons 
why the fluid thermalizes in about 1 fm/c, has shear viscosity to entropy density ratio $\eta/s$ close to its proposed lower bound of $1/4\pi$, and why such a system, about 10 fm across, can be described with hydrodynamics. Because of the small size of the fluid, thermal fluctuations in the fluid should contribute significantly to two-particle correlations and event-by-event fluctuations. This requires extending the well-known results of \cite{Landau:1980st} to relativistic hydrodynamics, preferably in the Landau frame 
\cite{Kapusta:2011gt}. Measuring the effect of thermal fluctuations can provide an important independent measurement of transport coefficients.

Thermal noise in the energy-momentum tensor has an autocorrelation function proportional to $\delta^4(x-x^\prime)$, which makes the variance of the cell-averaged energy 
and momentum densities proportional to $1/(\Delta V\Delta t)$. As a result, no matter how small the viscosity, there exists a minimum spatial grid size below which the results of non-perturbative, thermally fluctuating hydrodynamic simulations using white noise are unreliable and plagued by pathologies such as negative energy densities and gradients so large as to negate the application of hydrodynamics. This limit is not just a numerical artifact but is related to the coarse graining implicit in hydrodynamics.  Examination of thermal noise in \cite{Young:2013fka} showed how the transport coefficients themselves encode the limit of resolution of hydrodynamics;  including thermal noise in the most straightforward way has a limiting resolution built into it, unlike the algorithms without noise.  Viscous hydrodynamical codes for heavy-ion collisions even without thermal noise will have, for very limited times and spatial extents, viscous corrections that lead to unphysical pressures in the ideal part of the energy-momentum tensor $T^{\mu \nu}$. This problem is fixed in various {\it ad hoc} ways which all depend on the fact that the unphysical pressures occur only briefly and only at the earliest times of the collision or in the cold regions of the collisions beyond the freeze-out surface, where hydrodynamics is 
neither used nor expected to work. With the introduction of the noise term, these unphysical pressures occur more often, and these {\it ad hoc} methods are called more frequently to the point where one should be more skeptical of the accuracy of the results of these codes.

There appear to be two options to address the aforementioned difficulty.  If thermal fluctuations are to be included self-consistently and non-perturbatively in the space-time evolution of the system, it is probably necessary to use colored rather than white noise.  White noise correlators are proportional to Dirac delta-functions in space and time.  Their use is justified if the grid cells are larger than or at least comparable to the correlation length.  At sufficiently fine resolutions, the noise should actually be colored and correlated across cells.  Colored noise for the energy-momentum tensor appropriate for the matter created in heavy ion collisions has not been worked out, although for the baryon current it has been \cite{baryon-noise}.  The other option is to treat the noise as a perturbation on a noise-less background.  That is the approach first implemented in the Bjorken 1+1 dimensional fluid model in \cite{Kapusta:2011gt}.  It ought to be an accurate description of thermal fluctuations unless the equation of state has some critical behavior where fluctuations would be greatly amplified and carry the system to states far away from the average one; see, for example, \cite{Joe-Juan}.

In this paper we will use the 3+1 dimensional second-order viscous code \textsc{music} \cite{Schenke:2010nt,Schenke:2010rr} and treat thermal fluctuations perturbatively.  
A similar effort is discussed in \cite{Murase:2013tma}. We will apply it to Pb-Pb collisions at energies available at the LHC (Large Hadron Collider).  We verify that single-particle distributions are not affected by the implementation of noise.  Multi-particle distributions are affected measurably and this is demonstrated by a shift and increase in width of the momentum eccentricity distribution.

\section{Linearized Relativistic Hydrodynamics}
\label{linearized}

A robust method for simulating thermal noise must be both conservative, keeping $\partial_\mu T^{\mu \nu}_{{\rm tot}}=0$ where $T^{\mu \nu}_{{\rm tot}}$ is the full energy momentum tensor, and be able to handle discontinuities in the thermodynamic variables.  The energy-momentum tensor is separated into an averaged part $T^{\mu \nu}_0$, the fluctuating part of the ideal energy-momentum tensor $\delta T^{\mu \nu}_{{\rm id}}$ which arises from fluctuations in the local flow velocity and energy density, a similar fluctuating part of the tensor $\delta W^{\mu \nu}$ (in the Israel-Stewart notation), and the noise part $\Xi^{\mu \nu}$:
\begin{equation}
T^{\mu \nu}_{ {\rm tot} } = T^{\mu \nu}_0+\delta T^{\mu \nu}_{{\rm id}}+\delta W^{\mu \nu}+\Xi^{\mu \nu} {\rm .}
\end{equation} 
In linear response, the fluctuating parts and the noise part of $T^{\mu \nu}_{{\rm tot}}$ can be separated into a set of coupled equations. One can define $\delta W^\prime \equiv \delta W + \Xi$ to make one single 
stochastic partial differential equation to describe the sum of these two terms. The averaged part of the energy-momentum tensor is calculated using the Israel-Stewart 
equations:
\begin{equation}
\partial_\mu T^{\mu\nu}_{{\rm id.}}=-\partial_\mu W^{\mu\nu}{\rm ,}\nonumber
\end{equation}
\begin{equation}
\Delta^\mu_\alpha \Delta^\nu_\beta (u\cdot\partial) W^{\alpha\beta}=-\frac{1}{\tau}(W^{\mu\nu}-S^{\mu\nu})-\frac{4}{3}(\partial\cdot u)W^{\mu\nu}{\rm ,}\nonumber
\end{equation}
which are closed by the equation of state and the condition in the Landau frame that $u_\mu W^{\mu\nu}=0$, where 
$\Delta^{\mu\nu}=g^{\mu\nu}-u^\mu u^\nu$, $\Delta^\mu = \Delta^{\mu\nu}\partial_\nu$, and where the first order viscous correction in the Landau-Lifshitz frame is 
$S^{\mu\nu}=\eta(\Delta^\mu u^\nu+\Delta^\nu u^\mu -{\textstyle \frac{2}{3}}(\partial\cdot u)\Delta^{\mu\nu})$.
Similar to the Israel-Stewart equations for the averaged energy-momentum tensor, two coupled equations can describe the fluctuations. The first equation is as always just the law of conservation of energy and momentum:
\begin{equation}
\partial_t \delta T^{t \nu}_{{\rm id}} = -\partial_i \delta T^{i \nu}_{{\rm id}} - \partial_\mu \delta W^{\prime \mu \nu} {\rm ,}
\end{equation}
where $\delta W^{\prime\mu\nu}$ is the fluctuating part of the viscous correction to the energy-momentum tensor.

Up to first order in derivatives, the viscous correction is determined with 
a simple equation: 
\begin{eqnarray}
\delta W^{\mu\nu}_{(1)} \equiv \delta S^{\mu \nu} &=& \eta \big( \Delta^\mu \delta u^\nu + \Delta^\nu \delta u^\mu-{\textstyle \frac{2}{3}}(\partial \cdot \delta u)
\Delta^{\mu \nu}\big) \nonumber \\
&+& \delta \eta( \Delta^\mu u^\nu + \Delta^\nu u^\mu-{\textstyle \frac{2}{3}}(\partial \cdot u)\Delta^{\mu \nu}) \nonumber \\
&+& \eta (\delta\Delta^\mu u^\nu + \delta\Delta^\nu u^\mu -{\textstyle \frac{2}{3}}(\partial \cdot u)\delta\Delta^{\mu\nu}){\rm ,}\\
\nonumber
\end{eqnarray}
where the variations $\delta\Delta^{\mu\nu}$ and $\delta\Delta^\mu$ are up to 
first order in $\delta u^\mu$. One can confirm that this simple variation of $S^{\mu\nu}$ satisfies the Landau-Lifshitz condition at linear order:
\begin{equation}
(u+\delta u)_\mu (S+\delta S)^{\mu\nu} = u_\mu S^{\mu\nu} + u_\mu \delta S^{\mu\nu} + \delta u_\mu S^{\mu\nu} = 0{\rm .}
\end{equation}

The equation for $\partial_t \delta W^{\mu\nu}$ at second order in derivatives can be found in 
two ways: one is to start with the equation for $W^{\mu\nu}$,
\begin{equation}
(u\cdot \partial)W^{\mu\nu} = -\frac{1}{\tau_\pi}(W^{\mu\nu}-S^{\mu\nu})-\frac{4}{3}(\partial \cdot u)W^{\mu\nu}-u^\mu((u\cdot\partial)u_\alpha)W^{\alpha\nu}
-u^\nu((u\cdot\partial)u_\alpha)W^{\mu\alpha}{\rm ,}\nonumber
\end{equation}
examine its fluctuations and noise up to linear order, and include them in one equation:
\begin{eqnarray}
(u\cdot\partial)\delta W^{\prime \mu\nu} &=&  -\frac{1}{\tau_\pi}(\delta W^{\prime \mu\nu}-\delta S^{\mu\nu}-\xi^{\mu\nu})
-\frac{4}{3}(\partial \cdot \delta u)W^{\mu\nu}-\frac{4}{3}(\partial \cdot u)\delta W^{\prime\mu\nu}\nonumber \\
& & -\delta u^\mu((u\cdot\partial)u_\alpha)W^{\alpha\nu}-u^\mu(((\delta u\cdot\partial)u_\alpha)W^{\alpha\nu}+((u\cdot\partial)\delta u_\alpha)W^{\alpha\nu}
+(((u\cdot\partial)u_\alpha)\delta W^{\prime \alpha\nu}) \nonumber \\
& & -\delta u^\nu((u\cdot\partial)u_\alpha)W^{\alpha\mu}-u^\nu(((\delta u\cdot\partial)u_\alpha)W^{\alpha\mu}+((u\cdot\partial)\delta u_\alpha)W^{\alpha\mu}
+((u\cdot\partial)u_\alpha)\delta W^{\prime \alpha\mu}) \nonumber \\
& & -(\delta u\cdot\partial)W^{\mu\nu}{\rm .}  \nonumber \\
\label{IS}
\end{eqnarray}
Alternatively, starting with 
\begin{equation}
\delta\left[\Delta^\mu_\alpha \Delta^\nu_\beta (u\cdot\partial)W^{\alpha\beta}\right]=-\frac{1}{\tau_\pi}(\delta W^{\prime\mu\nu}-\delta S^{\mu\nu}-\Xi^{\mu\nu}){\rm ,}\nonumber
\end{equation}
(not a complete equation for $\delta W^\prime$ because $\Delta^\mu_\alpha$ is singular), a closed system of equations can be obtained by requiring 
the viscous part of $T^{\mu\nu}$ to be transverse:
\begin{equation}
(u+\delta u)_\mu (W+\delta W)^{\mu\nu} = u_\mu W^{\mu\nu} +  \delta u_\mu W^{\mu\nu} +  u_\mu \delta W^{\mu\nu} = 0{\rm ,}\nonumber
\end{equation}
which also yields Eq. \ref{IS}.
The noise term $\xi^{\mu \nu}$ has autocorrelation 
\begin{eqnarray}
\left \langle \xi^{\mu \nu}(x)\xi^{\alpha \beta}(x^\prime)\right \rangle &=& \bigg[2\eta T(\Delta^{\mu \alpha}\Delta^{\nu \beta}+\Delta^{\mu \beta}\Delta^{\nu \alpha}) \nonumber \\
& & +2(\zeta-2\eta/3)T\Delta^{\mu \nu}\Delta^{\alpha \beta} \bigg]\delta^4(x-x^\prime) {\rm ,}
\label{noise}
\end{eqnarray}
as discussed in \cite{Young:2013fka}.

When using Bjorken coordinates to describe ultrarelativistic heavy-ion collisions, the derivatives $\partial_\mu$ above must be replaced 
with their covariant counterparts $D_\mu$. The averaged part will be solved with the usual methods while the fluctuating and noise parts will be solved together, with the noise
acting as a source term.

The steps for finding solutions for $\delta T^{\mu \nu}$ and $\delta W^{\prime \mu \nu}$ are similar to those used for $T^{\mu \nu}_0$ and $W^{\mu \nu}_0$ in \cite{Schenke:2010nt}:
\begin{itemize}
\item Determine $\delta W^{\prime \mu \nu}$ at the next time step using the stochastic advective equation in Eq. \ref{IS}.
\item Next, determine $\delta T^{0 \nu}$ at the next time-step using $D_0 \delta T^{0 \nu} = -D_i \delta T^{i \nu}-D_\mu \delta W^{\prime \mu \nu}$. The numerical method should be conservative; the change in $\delta W^{\prime\mu\nu}$ during this timestep should be used to calculate $\partial_0 \delta W^{\prime\mu\nu}$.
\item Finally, determine $\delta T^{ij}$, as well as $\delta p$ and $\delta u^i$, using 
\begin{equation}
\delta T^{\mu \nu}_{ {\rm id} } = -\delta p g^{\mu \nu} + \delta p \left( 1+  (\dedp )_{n/s} \right) u^{\mu}_0 u^{\nu}_0 
+ (e_0+p_0)(u^{\mu}_0 \delta u^{\nu}+u^{\nu}_0 \delta u^{\mu})
\end{equation}
and a root-finding algorithm using the values of $\delta T^{0 \nu}$.

\end{itemize}

For the first step, the MacCormack method is a predictor-corrector method which alternates between upstream and downstream differencing:
\begin{eqnarray}
\bar{T}^{t \mu\; l}_i &=& T^{t \mu\; l}_i-\frac{T^{x \mu\; l}_{i+1}-T^{x \mu\; l}_i}{\Delta x}\Delta t{\rm ,} \nonumber \\
T^{t \mu\; l+1}_i &=& \frac{T^{t \mu\; l}_i+\bar{T}^{t \mu\; l}_i}{2}-\frac{\bar{T}^{x \mu\; l}_i-\bar{T}^{x \mu\; l}_{i-1}}{2\Delta x}\Delta t{\rm ,} \\
\end{eqnarray}
where $T^{t \mu\; l}_i$ is $T^{t\mu}$ averaged in the $i-$th cell at the beginning of the $l$-th timestep.
This is written in one dimension; for three dimensions, this is iterated for each direction. As a consequence of Godunov's theorem, the MacCormack method is more appropriate here than higher-order schemes because of the large gradients that are encountered in thermally fluctuating hydrodynamics.

\section{Thermal Fluctuations in Heavy-Ion Collisions}

Heavy-ion collisions are approximated fairly well as {\it boost-invariant} (only a function of $\tau$ and not $\eta$) when $\sqrt{s_{NN}}$ exceeds a few GeV. For this reason, it is advantageous to use $\tau$-$\eta$ coordinates instead of $t$, $z$, even if the hydrodynamic model is not boost-invariant. The space-time of the heavy-ion collision is still flat, but the Christoffel symbols for covariant derivatives are now non-zero:
\ba
D_\eta u^\tau &=& \partial_\eta u^\tau+\frac{1}{\tau}u^\eta{\rm ,} \nonumber \\
D_\eta u^\eta &=& \partial_\eta u^\eta+\frac{1}{\tau}u^\tau {\rm .}
\ea
The derivatives of any tensor can be determined by examining the derivatives of products of vectors. The derivatives must be modified for Bjorken coordinates.

The equation of state and transport coefficients for matter with temperatures above 120 MeV are highly non-trivial and are the focus of a continuing debate among lattice QCD practitioners and other nuclear physicists. These are some of the primary reasons for studying heavy-ion collisions. Finding out how observables are sensitive to the equation of state and transport coefficients, and how to infer them from data, should be the goal of hydrodynamical simulations. For now, we use one equation of state, determined by lattice QCD calculations \cite{Huovinen:2009yb}.  We also use temperature-independent values for $\eta/s$; the relaxation time $\tau_{\pi}$ varies slowly in \textsc{music} as $1/T$ and 
for now, we approximate it as being constant in the calculation of the fluctuations.  Future work where other sources of event-by-event fluctuations are also included will use a variety of values for viscosity, relaxation time, and equation of state with the goal of using heavy-ion collisions and simulation together to determine the properties of hot nuclear matter.

The final hadronic observables are measured after the freeze-out of the flowing matter into freely streaming hadrons. The effect of non-equilibrium corrections to the energy-momentum tensor is the subject of research itself \cite{Dusling:2009df}. We extend the correction to freeze-out spectra discussed in \cite{Teaney:2002zt} to include the contribution from thermal noise:
\begin{equation}
\delta f \propto f_0(1\pm f_0) \left(W_{\alpha \beta}+\delta T_{\alpha \beta}+\delta W^\prime_{\alpha \beta}\right)\frac{p^\alpha p^\beta}{2(e_0+p_0)T^2} {\rm ,}
\end{equation}
where the +($-$) comes from quantum statistics of bosons (fermions). The freeze-out is isothermal, and below this temperature the noise is also set to zero.

\section{Results at the LHC}

\begin{figure}[ht]
\centering
\includegraphics[trim = 25mm 5mm 20mm 10mm, clip,width=8cm]{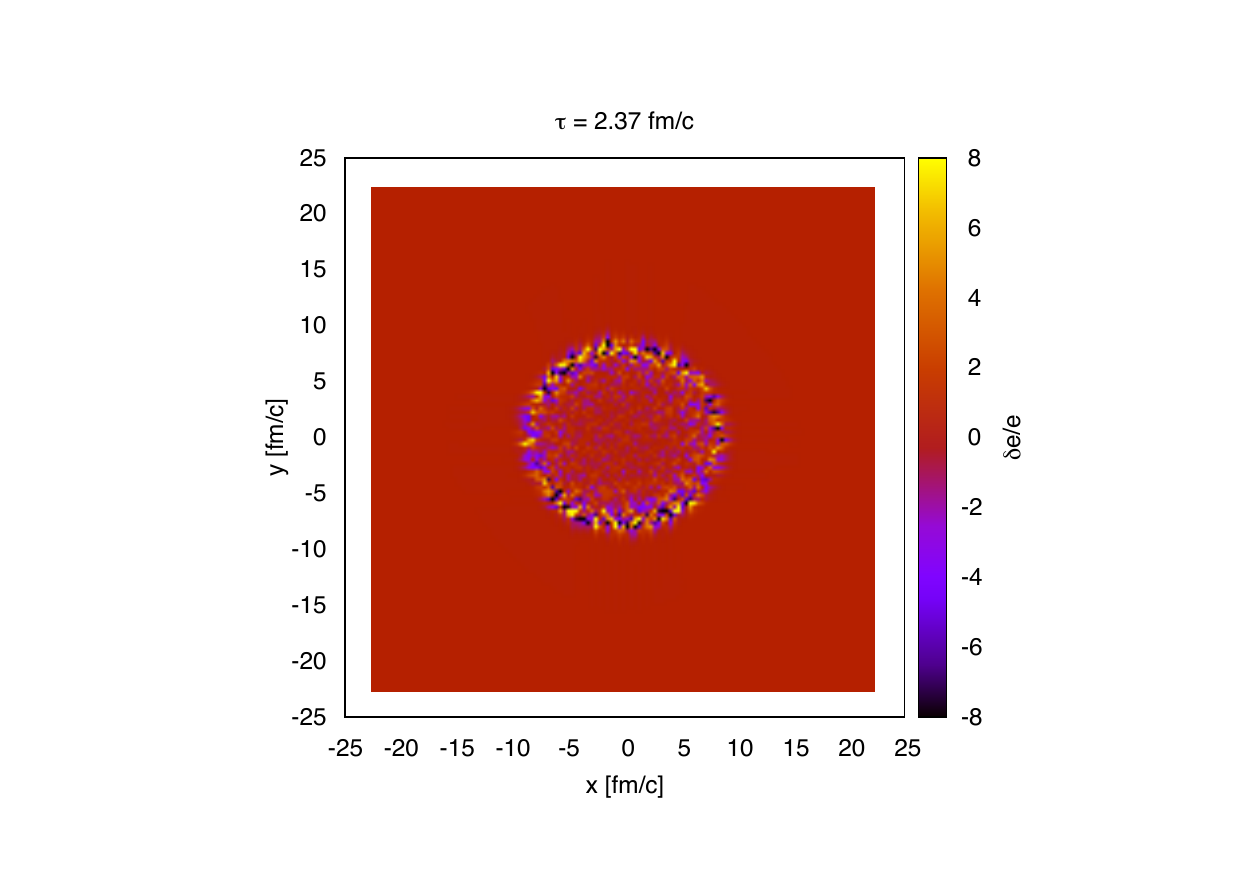}
\includegraphics[trim = 25mm 5mm 20mm 10mm, clip,width=8cm]{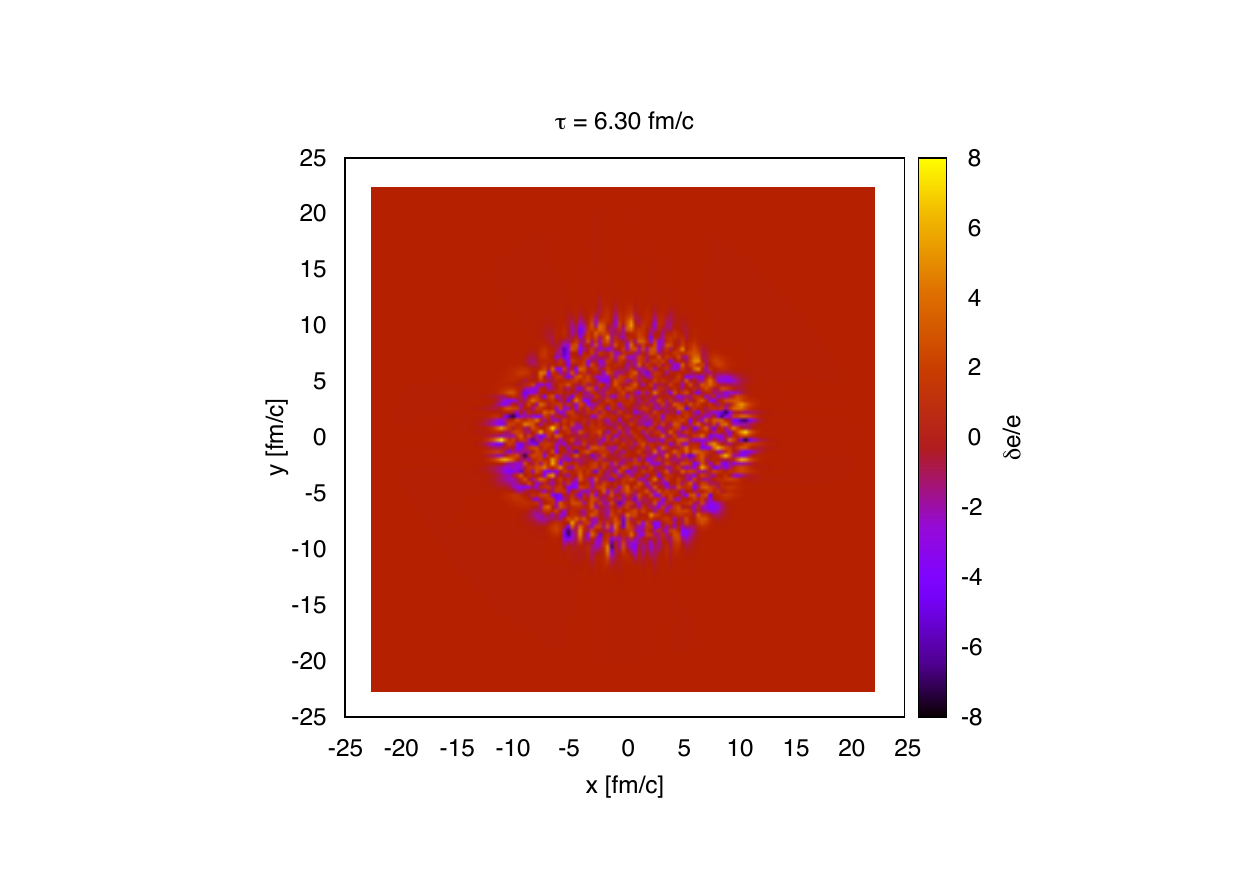}
\includegraphics[trim = 25mm 5mm 20mm 10mm, clip,width=8cm]{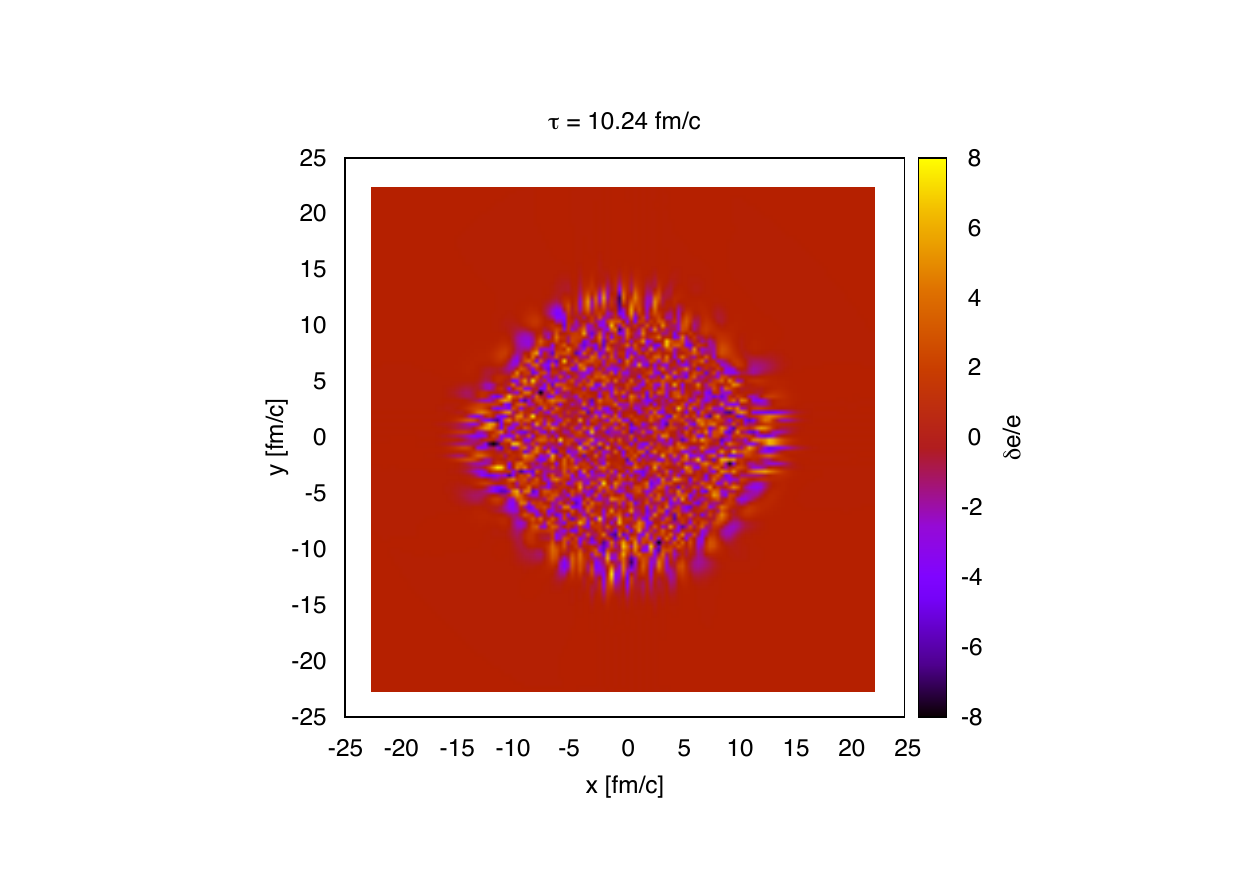}
\includegraphics[trim = 25mm 5mm 20mm 10mm, clip,width=8cm]{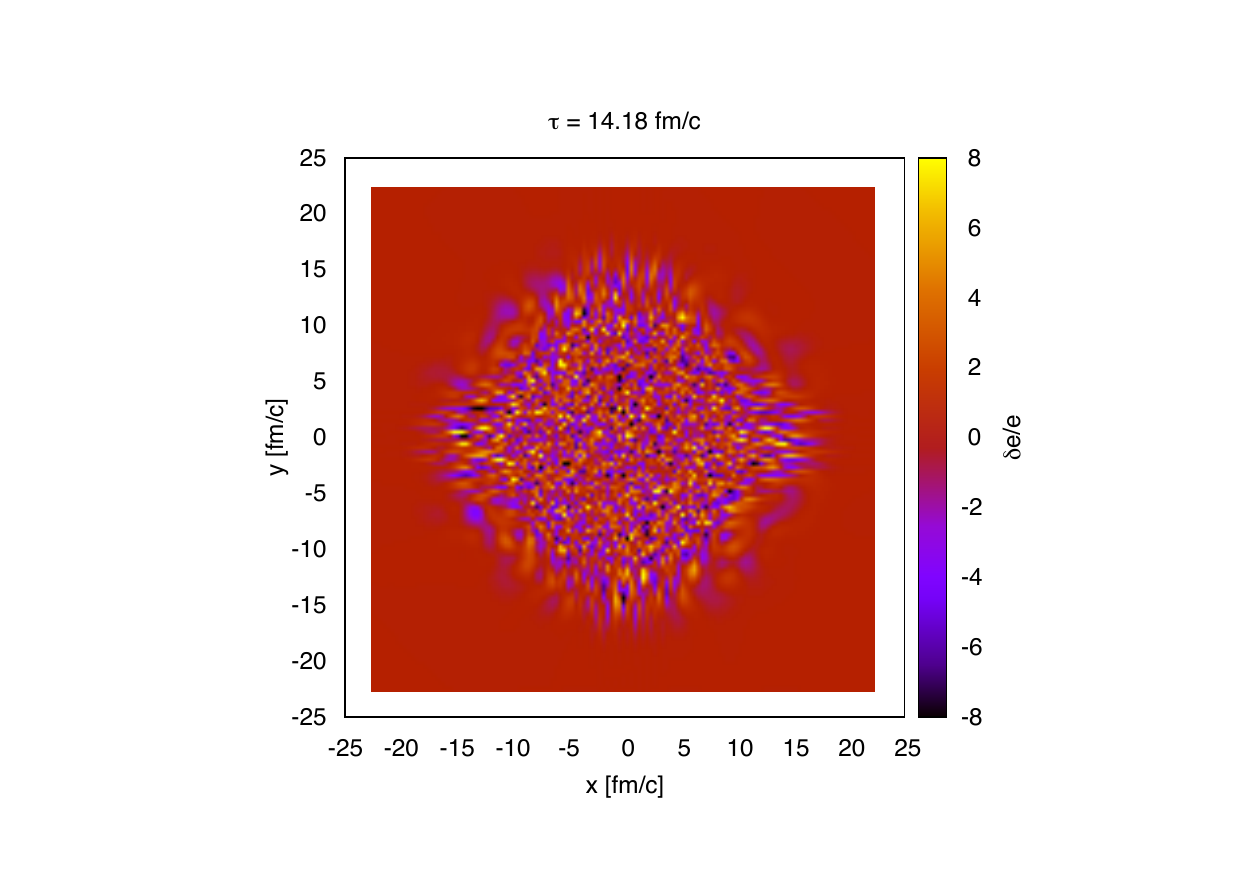}
\caption[]{Fluctuations in energy density in the transverse plane as a function of time in a typical Pb+Pb collision for $b = 6$ fm at LHC energies. (color online)}
\label{deltaE}
\end{figure}

With these modifications, a set of 200 thermally fluctuating events with impact parameter $b = 6.0$ fm at the LHC are calculated using \textsc{music}.  The underlying event 
is initialized with smooth initial conditions and with all fluctuating parts of the energy momentum tensor set to $0$. The impact parameter corresponds to a typical event in the 10-20\% centrality class. The cells have transverse dimensions of $\Delta x = \Delta y = 45/128\;{\rm fm} \approx 0.35$ fm.  The cells begin evolving at $\tau_i=0.4$ fm/c and then all cells are frozen out by about $\tau_f = 14.9$ fm/c.  Figure \ref{deltaE} 
shows the evolution of $\delta e/e$, the ratio of the thermal fluctuation in local energy density to its average value. 


\begin{figure}[h]
\centering
\includegraphics[width=0.69\textwidth]{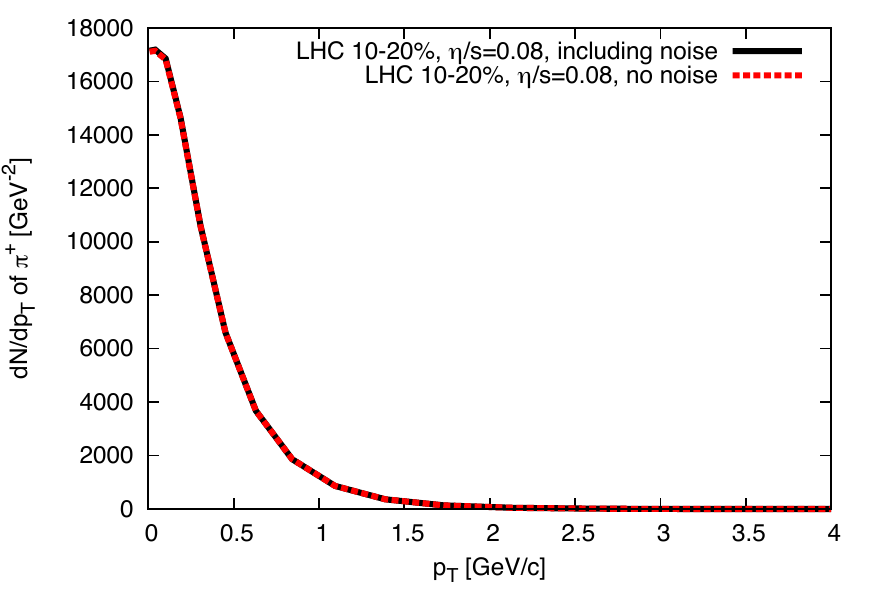}
\includegraphics[width=0.69\textwidth]{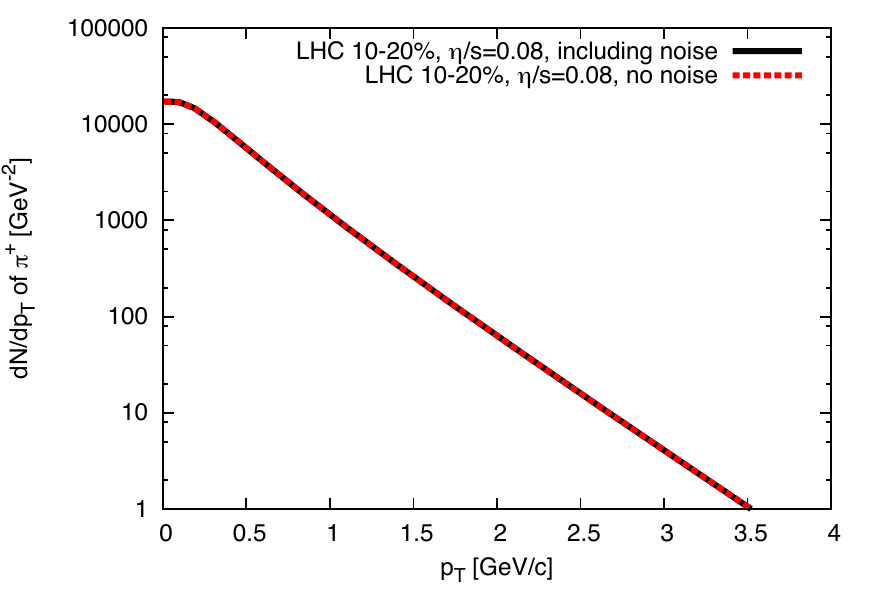}
\caption[]{Transverse momentum distribution for pions in linear and log scales showing that the single particle distribution is unaffected by noise.  (color online)}
\label{dNdpT}
\end{figure}

Figure \ref{dNdpT} shows the distributions of $\pi^+$ mesons for both the case without fluctuations and for the average of 200 thermally fluctuating events. Because 
$\left\langle \delta T^{\mu \nu} \right\rangle = \left\langle \delta W^{\prime \mu \nu} \right\rangle = 0$, the average of a large ensemble of thermally fluctuating events should have no effect on $\left\langle dN/dp_T \right\rangle$, which Figure \ref{dNdpT} demonstrates. This is one important test of the methods of Section \ref{linearized}.

However, thermal noise affects not only the average value of the harmonic coefficient $v_n$ but also leads to event-by-event variations in $v_n$ at the same impact parameter.  Thermal noise tends to increase average values of $v_n$. To see this, imagine a collision with zero impact parameter. In the averaged case, this leads to a cylindrically symmetrical expansion, and $v_n=0$ for all $n = 0$. However in a single event with noise, $v_n \neq 0$ in general. Because each $v_n$ is defined to be positive, $\left\langle v_n \right\rangle \neq 0$ when thermal noise is included.


\begin{figure}[h]
\centering
\includegraphics[width=0.69\textwidth]{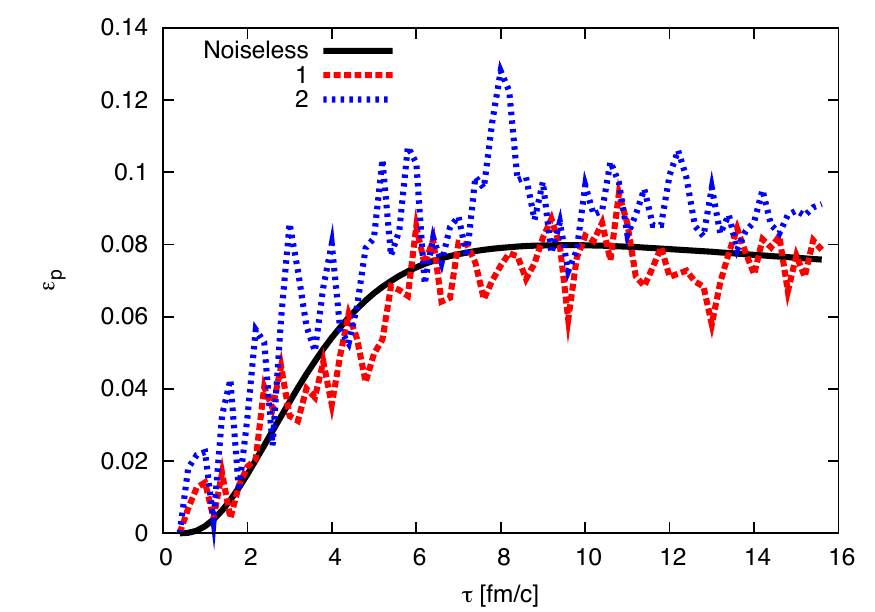}
\caption[]{The evolution in time of $\epsilon_p$ from the calculations of two $b = 6$ fm collisions with thermal noise as well as from a calculation without noise. Notice 
both the rapidly decorrelating variations in $\epsilon_p$, coming from $\delta W^{\prime \mu \nu}$, as well as the variations over longer time-scales.  (color online)}
\label{eP}
\end{figure}


\begin{figure}[h]
\centering
\includegraphics[width=0.69\textwidth]{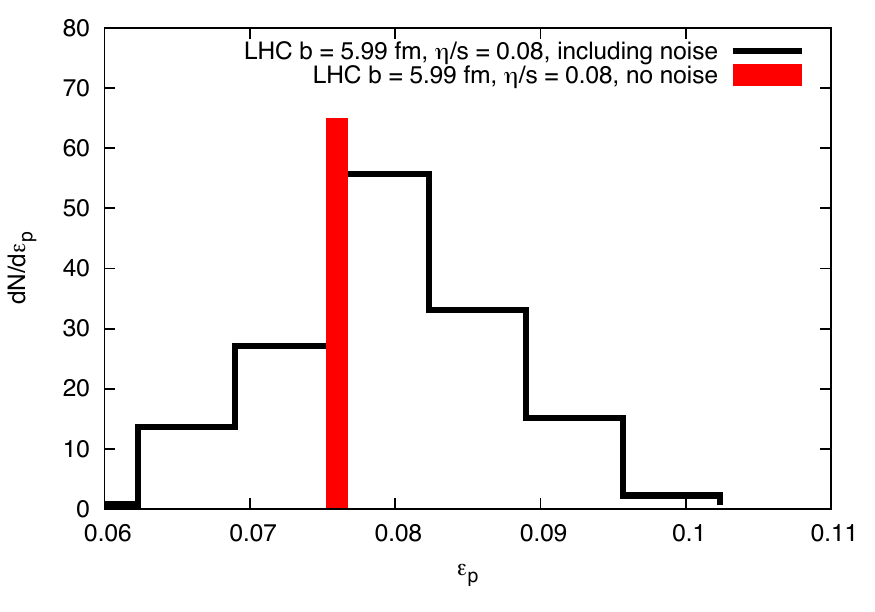}
\includegraphics[width=0.69\textwidth]{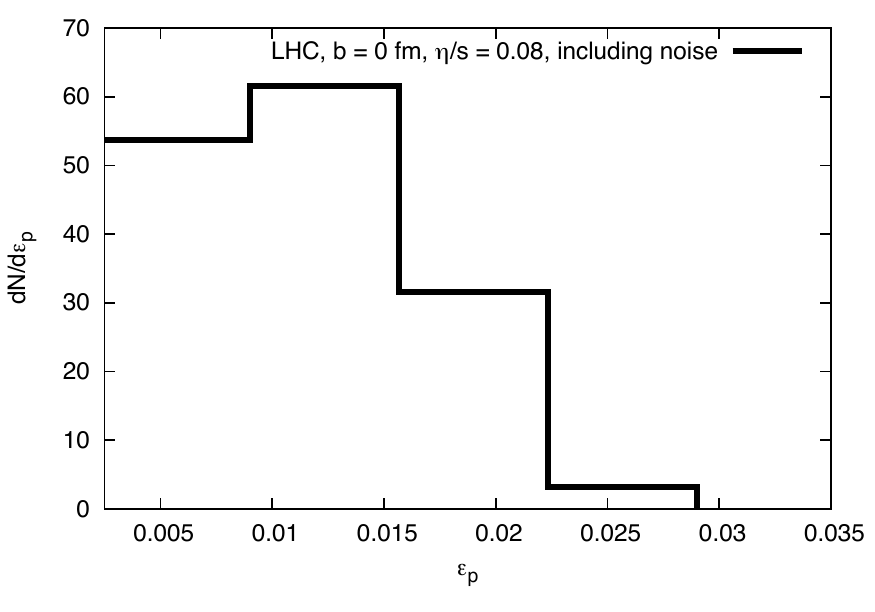}
\caption[]{Top panel ($b = 6$ fm): Inclusion of noise increases the average $v_2$ and broadens its distribution.  Bottom panel ($b = 0$ fm): Even for exactly central collisions there is a nonzero average $v_2$ due to noise. (color online)}
\label{dNdv2}
\end{figure}

For now, to avoid the extra complications of freeze-out to individual hadrons and viscous corrections to thermal distribution functions, we calculate the momentum eccentricity 
\be
\epsilon_p = \sqrt{ \frac{\langle T^{xx}-T^{yy} \rangle^2 + \langle 2T^{xy} \rangle^2}{\langle T^{xx} + T^{yy} \rangle^2} } {\rm ,}
\ee
which is a generalization of 
the quantity discussed in \cite{Ollitrault:1993ba}.  It is a proxy for $v_2$.  Here $T^{\mu \nu}$ represents the total energy-momentum tensor, including viscous corrections as well as noise, which gives the best approximation to $\sum_i \frac{p^\mu_i p^\nu_i}{E_i} \delta^3({\bf x}-{\bf x}_i(t))$. Figure \ref{eP} shows $\epsilon_p$ for two thermally fluctuating hydrodynamical events as functions of proper time. 

The top panel of Figure \ref{dNdv2} shows the probability distribution for events with impact parameter of $b=6.0$ fm at the LHC.  Without noise there is of course just one value.  With noise there is a modest broadening of the distribution as well as an increase in the average value by about 5\%.  The bottom panel of figure \ref{dNdv2} shows the distribution of $\epsilon_p$ for an ``ultra-central" event with  impact parameter $b= 0$ fm. Interestingly, $\langle \epsilon_p \rangle$ is non-vanishing thanks to thermal noise driving all $\langle v_n \rangle$ to non-zero values, as argued above.

\section{Conclusions}

In this paper thermal noise has been included in second-order viscous hydrodynamics by an extension of the Israel-Stewart formalism. After some special 
considerations for the fluid produced in heavy ion collisions, the effect of these fluctuations was calculated for Pb-Pb collisions at the LHC. A small variance was found in 
$\epsilon_p$, the momentum eccentricity, for impact parameter $b = 6$ fm, while a significant variance was found for $b = 0$.

Thermal noise must contribute to event by event fluctuations thanks to the fluctuation-dissipation theorem; however, our results confirm that the contribution is often subleading when compared with results sampling initial-state fluctuations. Rather, the significance of thermal noise lies in its connection to transport coefficients: the variances caused by 
thermal noise are proportional to the transport coefficients, possibly allowing for a measurement of the shear viscosity in heavy ion collisions independent of the previous 
determinations based on elliptic flow. In \cite{Aad:2013xma}, the ATLAS collaboration presented results for distributions of flow harmonics in several event classes; the variances of $v_2$ in the mid-central classes are approximately 0.05.  The variances are explained well in many centrality classes with the methods employed in \cite{Schenke:2013aza}. However, note the results from the ``ultra-central" CMS 0-0.2\% centrality class \cite{CMS:2012tba}.   The average values and variances of $v_n$ are on the order of our results from only thermally fluctuating hydrodynamics; the results from other calculations only explain some of the integrated $v_n$  with unusually large values for $\eta/s$. A calculation combining initial state fluctuations and thermal fluctuations, aimed at describing the ultra-central event class,  might not only explain this data but also provide another measurement of $\eta/s$.

Additional work in both the theory and simulation of thermal noise will improve the understanding of flow in heavy ion collisions. One important direction for theoretical 
improvements is to go beyond linear response: the quality of any truncated perturbative expansion is most reliably estimated with a full calculation at the next order. Recently, 
this work has become easier with the proposal of an effective action for hydrodynamics \cite{Kovtun:2014hpa}. Finally, even linearized hydrodynamics poses problems for 
numerical simulation: the gradients become larger with decreasing cell size, seeming to rule out the possibility of using any higher-order method. In this paper we used the 
MacCormack method for its ability to perform well in the presence of shocks. However, a detailed investigation into the performance of numerical methods for linearized 
fluctuating hydrodynamics would be useful.  Such work is underway.

\section{Acknowledgments}

JK and CY are supported by the U.S. DOE Grant No. DE-FG02-87ER40328. 
CG and SJ are supported by funding from the Natural Sciences and Engineering Research Council of Canada. 
BPS is supported under DOE Contract. No. DE- AC02-98CH10886. 
CY thanks E. Shuryak for very helpful suggestions. 
We are grateful for resources from the University of Minnesota Supercomputing Institute.

\appendix

%


\begin{thebibliography}{9}

\bibitem{Adcox:2004mh} 
  K.~Adcox {\it et al.}  [PHENIX Collaboration],
  Nucl.\ Phys.\ A {\bf 757}, 184 (2005).

\bibitem{Adams:2005dq} 
  J.~Adams {\it et al.}  [STAR Collaboration],
  Nucl.\ Phys.\ A {\bf 757}, 102 (2005).

\bibitem{Landau:1980st}
   L.~D.~Landau and E.~M.~Lifshitz, {\it Statistical Physics: Part 2}
   (Pergamon, Oxford, 1980).

\bibitem{Kapusta:2011gt} 
  J. I. Kapusta, B. Muller and M. Stephanov, Phys. Rev. C {\bf 85}, 054906 (2012).

\bibitem{Young:2013fka} 
  C. Young, preprint arXiv:1306.0472.
 
\bibitem{baryon-noise}
J. I. Kapusta and C. Young, preprint arXiv:1404.4894.

\bibitem{Joe-Juan}
J. I. Kapusta and J. M. Torres-Rincon, Phys. Rev. C {\bf 86}, 054911 (2012).

\bibitem{Schenke:2010nt} 
  B. Schenke, S. Jeon and C. Gale, Phys. Rev. C {\bf 82}, 014903 (2010).
  
\bibitem{Murase:2013tma} 
  K.~Murase and T.~Hirano,
  arXiv:1304.3243.

\bibitem{Schenke:2010rr} 
  B. Schenke, S. Jeon and C. Gale, Phys. Rev. Lett. {\bf 106}, 042301 (2011).

\bibitem{Huovinen:2009yb} 
  P. Huovinen and P. Petreczky, Nucl. Phys. {\bf A837}, 26 (2010).

\bibitem{Dusling:2009df} 
  K. Dusling, G. D. Moore and D. Teaney, Phys. Rev. C {\bf 81}, 034907 (2010).

\bibitem{Teaney:2002zt} 
  D. Teaney, Nucl. Phys. {\bf A715}, 817 (2003).
  
\bibitem{Ollitrault:1993ba} 
  J.-Y. Ollitrault, Phys. Rev. D {\bf 48}, 1132 (1993).
 
\bibitem{Aad:2013xma} 
  G. Aad {\it et al.}  [ATLAS Collaboration], JHEP {\bf 1311}, 183 (2013).

\bibitem{Schenke:2013aza} 
  B. Schenke, P. Tribedy and R. Venugopalan, preprint  arXiv:1312.5588.
 
\bibitem{CMS:2012tba} 
  CMS Collaboration [CMS Collaboration],
  CMS-PAS-HIN-12-011.
 
\bibitem{Kovtun:2014hpa} 
  P. Kovtun, G. D. Moore and P. Romatschke, preprint  arXiv:1405.3967.


\end{thebibliography}
\end{document}